\documentclass[%
 reprint,
 amsmath,amssymb,
 aps,
]{revtex4-2}

\usepackage{graphicx}% Include figure files
\usepackage{dcolumn}% Align table columns on decimal point
\usepackage{bm}% bold math
\usepackage{float}
\usepackage{color}
\usepackage{epstopdf}
\usepackage{tikz}
\usepackage{amsmath}
\usepackage[toc]{appendix}
\usepackage{graphicx,epstopdf}
\DeclareUnicodeCharacter{0301}{*************************************}
\newcommand{\tikzcircle}[2][red,fill=red]{\tikz[baseline=-0.5ex]\draw[#1,radius=#2] (0,0) circle ;}%
\usepackage{hyperref}% add hypertext capabilities

\begin{document}

\preprint{APS/123-QED}

\title{Dynamics of colloidal rods rotating in viscoelastic media}% Force 

\author{N Narinder$^{1,3}$}
 \author{Jyotiprakash Behera$^2$}
\author{Ambarish Ghosh$^{1,2}$}%
 %\email{clemens.bechinger@uni-konstanz.de}
\affiliation{%
 1. Centre for Nano Science and Engineering, Indian Institute of Science, Bengaluru, India\\%
 2. Department of Physics, Indian Institute of Science, Benglauru, India\\
 3. Physics of Life, Technische Universit\"{a}t Dresden, Germany}

% \begin{abstract}
% We experimentally investigate the in-plane rotational motion of ferromagnetic colloidal rods in viscoelastic media under a rotating magnetic field. Contrary to their rotation in a Newtonian fluid, in a viscoelastic fluid the rods continue to exhibit a net angular drift even for applied field frequencies which are an order of magnitude larger than the step-out frequency. Despite experimental evidence, previous studies failed to explain the observed behavior. This is due to the inherent assumption that the rods' angular velocity beyond step-out follows same dependence on the applied field frequency as the Newtonian fluids. We demonstrated that the observed net rotation of rods after step-out originates from their interaction with the microstructural stress-relaxation processes within the viscoelastic fluid. Consequently, it exhibits a strong dependence on the rheological properties of the fluid. Our results are further supported by a minimal model which incorporates the memory-mediated response of the viscoelastic fluid on their motion. Furthermore, we demonstrate, both experimentally and numerically, that the observed effect represents a generic feature of viscoelastic media and is expected to manifest for elongated probes in other complex surroundings such as biological assays and colloidal glasses.

% \end{abstract}

\begin{abstract}
{We experimentally investigate the in-plane rotational motion of ferromagnetic colloidal rods immersed in viscoelastic media and subjected to a rotating magnetic field. Unexpectedly, we observe significant angular velocity even at field frequencies an order of magnitude exceeding the step-out frequency, a regime where rods typically cease rotating in Newtonian fluids. This anomalous behavior arises from the interplay between the rapid rod actuation driven by the external field and the slower microstructural relaxation of the viscoelastic fluid. A minimal model incorporating memory effects quantitatively captures our experimental findings. Our study demonstrate a rather general case of microrheological probe dynamics in viscoelastic media where the behavior beyond step-out frequency depends strongly on the rheological parameters medium. Additionally, we derive an analytical expression for the rod orientation in the high-frequency limit, providing a potential method for extracting rheological parameters.}
 
\end{abstract}

%\keywords{Suggested keywords}%Use show keys class option if keyword
                              \maketitle

%\tableofcontents
\section{Introduction}
Active microrheology has emerged as an versatile technique for the mechanical characterization of fluids and biological assays sample of volumes as small as $\sim 1\,\mu l$. In contrast to its passive counterpart, it employs an additional external perturbation for probing which leads to a significantly higher signal-to-noise ratio. Additionally, it offers to tailor the perturbation to specific amplitudes and frequencies, thus, encompassing investigations for broader range of rheological characterization. Within the expanding field of active microrheology, magnetic rotational rheometry has emerged as a particularly powerful micro-mechanical technique for characterizing biological systems \textit{e.g.}, living cells. This method utilizes the rotational response of strategically designed, biocompatible micro-probes to externally applied rotating magnetic fields~\cite{bausch1998local,cheng2003rotational,wilhelm2008out,zhang2012targeted,tokarev2012magnetic,berret2016local,wu2018swarm,radiom2021magnetic}. Unlike other techniques, these microprobes can be seamlessly integrated with living cells due to their biocompatibility, ensuring minimal disruption to the cell cycle and overall biological function.~\cite{venugopalan2020fantastic,zhou2021magnetically}. These attributes position such micro-probes as a highly promising tool for rheological characterization of biological samples. In Newtonian fluids, with increasing the frequency of the applied magnetic field $\omega_B$, such probes undergo a transition from synchronous rotation to an asyncrhonous slip-stick rotational motion with the applied external field~\cite{dhar2007orientations,coq2010three,ghosh2013analytical,chevry2013magnetic}. Mathematically, the rotational motion of the probes can be described as
 \begin{equation} \label{eq:cutoff_freq}
	\begin{split}
		\Omega & = \omega_B~\hspace{2.5cm}\textrm{for}~~\omega_B\leq\Omega_C\\
		\Omega &=\omega_B-\sqrt{\omega_B^2-\Omega_C^2}~\hspace{0.4cm}\textrm{for} ~~\omega_B\geq\Omega_C
	\end{split}
\end{equation}

where $\Omega_C$ is the onset frequency of the transition known as the step-out frequency which is given by the ratio of the applied magnetic force to the viscous friction as $\Omega_C=(\mathbf{|m||B|})/(\gamma\eta_0)$ where $\mathbf{m}$ and $\mathbf{B}$ denote the  magnetic moment of the probe and the applied magnetic field while $\gamma$ and $\eta_0$ are the rotational friction coefficient of the probe and the viscosity of the fluid. Eqn.~\ref{eq:cutoff_freq} beyond step-out \textit{i.e.}, $\omega_B\geq\Omega_C$ is valid when the characteristic relaxation timescales of the system are well separated from the activation timescale $\tau_{act}=2\pi/\omega_B$ set by the frequency of the external field $\omega_B$. The relaxation timescales include the viscous relaxation $\tau_{vis}=\gamma\eta_0/\mathbf{|m||B|}$ and the stress-relaxation time of the viscoelastic fluid $\tau_0$. We ask, how the rotational motion of the rod get modified in a more general scenario when the system lacks a clear-cut separation of these timescales? 
To address this question, we examined the in-plane rotational motion of colloidal rods within a viscoelastic micellar medium. The stress-relaxation times of this medium, carefully adjusted by varying micelle concentration, were tuned to be on the order of a few seconds while the activation timescales were lowered to $\tau_{act}\sim 0.01~\mathrm{s}$. This leads to an overlap of the viscoelastic relaxation process with the external activation of the  system, resulting in the rods dynamics strongly governed by the memory effects. As a consequence, we observed a significant angular motion of the rods ($\sim 0.2\,\Omega_C $) even at applied frequencies as large as $15\,\Omega_C$. At such high field frequencies, the rods in the counterpart Newtonian fluid would only rotate with $\Omega\approx0.03~\Omega_C$ which is an order of magnitude smaller compared to what we experimentally observed $0.2~\Omega_C$ in the viscoelastic case. Additionally, unlike the Newtonian case,  
the rotational frequency beyond $\Omega_C$, exhibits a pronounced dependence on the fluid's rheological properties. We explained that the observed net rotation arises from the interplay between their rotational motion and the relaxation processes of induced strains within the viscoelastic network. This proposed mechanism demonstrates an excellent agreement with a minimal non-Markovian model which explicitly incorporates the memory effects of the surrounding fluid. Our numerical simulations further corroborate this, revealing a direct correlation between the observed net rotation and the increasing stress-relaxation time of the fluid. Our findings, thus, hint towards the generic nature of the observed net rotation for systems with extended relaxation times.

\section{Experimental details}
\begin{figure}[!]
    \centering
    \includegraphics[scale=0.60]{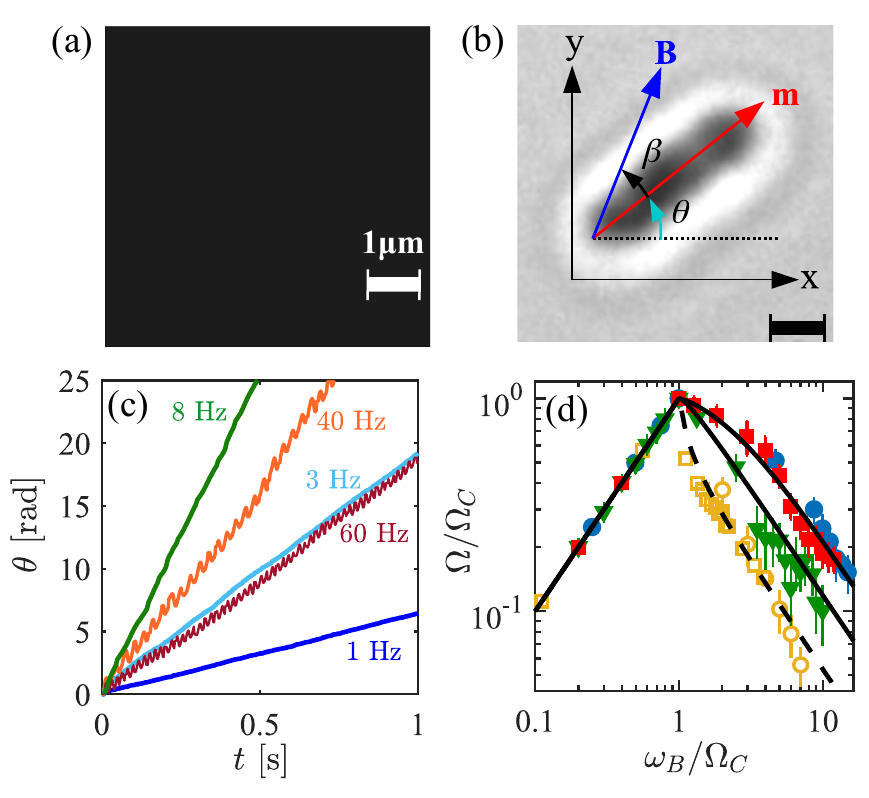}
    \caption{(a) An SEM image of the colloidal rod where the two brighter portions correspond to the $\mathrm{Fe-Co}$ layers. (b) Relevant coordinates to describe the 2D rotational motion of the rod. The scale bar is $2~\mathrm{\mu m}$. (c) Steady state time evolution of the angular coordinate $\theta$ of a rod rotating in $7~\mathrm{mM}$ CPyCl/NaSaI solution at $\mathbf{|B|=30~\mathrm{G}}$ for various applied field frequencies. (d) Dependence of the angular velocity $\Omega$, normalized by the corresponding step-out value $\Omega_C$, on the applied frequency of the external magnetic field $\omega_B$ for CPyCl/NaSal solution at $10~\mathrm{G}$ ($\tikzcircle[fill=black]{3pt}$), $30~\mathrm{G}$ ($\blacksquare$) and for CTAB/NaSal solution at $30~\mathrm{G}$ ($\blacktriangledown$). The corresponding results for the Newtonian (water-glycerol) solution are shown with the open symbols. The solid and dashed curves represent the corresponding numerical simulations.}
   \label{fig:panel1}
\end{figure}

The colloidal rods used in our experiments are grown on a Si wafer consisting of pillars with spacing $1~\mathrm{\mu m}$, using the glancing angle deposition technique. A detailed description of the process is mentioned in Ref.~\cite{brett2008new}. In comparison to the process for helices development~\cite{ghosh2009controlled,fischer2011magnetically}, here, we kept the rotation speed much smaller ($\approx 0.03~\mathrm{rad\,s^{-1}}$) which led to the growth of rod structures. The length $L$ and thickness $\sigma$ of the rods are $5~\mu\mathrm{m}$ and $1~\mu\mathrm{m}$, respectively. In order to impart a permanent magnetic moment to the rods, we alternatively sandwiched two iron-cobalt layers, each of thickness $150~\mathrm{nm}$ between the $\mathrm{SiO_2}$ during the growth process. The rods are then magnetized such that their magnetic moment vector orients parallel to their long axis. A SEM image of a rod is shown in Fig.~\ref{fig:panel1}\,(a). The suspension of the developed rods is then obtained in ultra-pure water by sonication of the wafer. As the viscoelastic fluid, we use an equimolar solution of surfactant, cetylpyridinium chloride monohydrate (CPyCl) and sodium salicylate (NaSaI) in water at a concentration of $7~\mathrm{mM}$. Under our experimental conditions, the fluid exhibits a viscoelastic nature~\cite{gomez2015transient,berner2018oscillating}. Using passive microrheology~\cite{mason1997particle}, the zero shear viscosity $\eta_0$, viscosity at infinite shear $\eta_\infty$, and the stress-relaxation time $\tau$ of the fluid are determined to  $0.25\pm 0.02~\mathrm{Pa\,s}$, $0.04\pm 0.01~\mathrm{Pa\,s}$ and $1.73\pm0.03~\mathrm{s}$, respectively (see Supplemental material for details). A small volume fraction of the rods suspended in the viscoelastic fluid is confined using a thin sample cell and placed between the two Helmholtz coils along the defined $x$ and $y$ axis. Owing magnetic moment along the long axis, the rods display in-plane rotational motion when subjected to a rotating magnetic field $\mathbf{B}$ along the $x-y$ plane. To study the distinct features of the rotational rod motion, the field frequencies $\omega_B$ are varied between $1-100~\mathrm{Hz}$ at field amplitudes of $|\mathbf{B}|=10~\mathrm{G}$ and $30~\mathrm{G}$. It should be noted for the applied $|\mathbf{B}|$ values, the magnetic moment remained parallel to the long axis of the rod, as verified separately by its response to the applied field direction. The rod rotational motion is described by the angular coordinates $\theta$  and $\beta$ \textit{i.e.}, the angle which long axis of the rod subtends with the $x-$axis and the instantaneous magnetic field vector $\mathbf{B}$, respectively, as depicted in Fig.~\ref{fig:panel1}\,(b).  

\section{Experimental results}
In Fig.~\ref{fig:panel1}\,(c), we plotted the typical time evolution of the rod angular coordinate $\theta(t)$ rotating in the viscoelastic CPyCl/NaSal solution at various frequencies $\omega_B$ of the applied field $\mathbf{|B|}=30~\mathrm{G}$. Clearly, at smaller 
$\omega_B$, $\theta(t)$ in the steady state ($t>>\eta_\infty\tau/\eta_0$) evolves linearly with its slope equals $\omega_B$, thus, marking a persistent synchronous rotation with $\mathbf{B}$. The synchronous angular motion results from the balance between the angular drag $\gamma_\theta\Omega$ and the magnetic torque $\mathbf{m\times B}=\mathbf{|m||B|}\sin\beta$ where $\beta$ acquires a constant value. Such behavior, however, is not expected to occur for all $\omega_B$~\cite{frka2011dynamics}. This is demonstrated by the angular trajectories corresponding to $\omega_B=40~\mathrm{Hz}$ and $60~\mathrm{Hz}$, where, $\theta(t)$ undergoes a slip-stick motion with the field. In this regime, $\mathbf{m}$ tend to rotate along $\mathbf{B}$ for certain time \textit{i.e.}, stick phase and eventually slips and rotates opposite to $\mathbf{B}$ \textit{i.e.}, slip phase. Consequently, the rod rotates with a net reduced mean angular velocity $\Omega$ with respect to the field as evident from the slopes corresponding to the curves at $40~\mathrm{Hz}$ and $60~\mathrm{Hz}$. In Fig.~\ref{fig:panel1}\,(d), we plot the measured slopes \textit{i.e.}, $\Omega$ as a function of $\omega_{B}$ for the viscoelastic fluid 
at $10~\mathrm{G}$ (solid circle) and $30~\mathrm{G}$ (solid square). The  measurements corresponding to the Newtonian case are shown with similar open symbols. In agreement with the previous studies ~\cite{ghosh2012dynamical,ghosh2013analytical}, mean $\Omega$ in a purely viscous liquid can be well described by Eqn.~\ref{eq:cutoff_freq} which is plotted as dashed curve in Fig.~\ref{fig:panel1}\,(d). A slight reduction in measured $\Omega/\Omega_C$ compared to expectation of Eqn.~\ref{eq:cutoff_freq} is possibly due to hydrodynamic interaction with the sample cell walls which effectively increases the viscous friction.\\

On the other hand, Eqn.~\ref{eq:cutoff_freq}, is not general and failed to describe the behavior of the rods in our viscoelastic cases where the characteristic relaxation times $\eta_\infty\tau/\eta_0$ are longer than the Newtonian case by orders of magnitude. Evidently, unlike the Newtonian case, here, $\Omega$ drops much slower as a function of $\omega_{B}$ and remains substantially finite ($\Omega/\Omega_C\approx0.2$) even for field frequencies as high as $15\,\Omega_C$. This is in significantly differnt from the expectation in a  Newtonian case where at $15\,\Omega_C$, $\Omega/\Omega_C$ is expected to be $\approx0.03$ which is vanishingly small, as found in our experiments. In contrast to the  $\omega_B-$dependence, its dependence on the applied field amplitude $\mathbf{|B|}$ is linear for both fluids. This is illustrated in Fig.~\ref{fig:panel1}\,(d), where the normalization of the measured $\Omega$ by the step-out frequency for different $\mathbf{|B|}$ converges to a single curve for both the cases.\\

To confirm whether the observed net rotation of rods even at such large frequencies is a feature specific to CPyCl/NaSaI solution, we also performed experiments with equimolar cetyltrimethyl ammonium bromide (CTAB) and sodium salicylate (NaSaI) in water at concentration $7~\mathrm{mM}$~\cite{inoue2005nonlinear} whose rheological parameters are significantly different from CPyCl/NaSaI solution (see Supplemental material). The obtained results are plotted, as solid triangles, in Fig.~\ref{fig:panel1}\,(e). Similar to the previous micellar solution, a net $\Omega$ is observed even for the applied frequencies an order higher than the step-out value $\Omega_\mathrm{C}$ which confirms that the effect is rather generic feature of such fluids. Notably, the behavior beyond step-out is markedly different from the CPyCl/NaSaI case. This hints at its strong dependency on the rheological properties of the fluid. 

\begin{figure*}[!]
	\centering
	\includegraphics[scale=0.44]{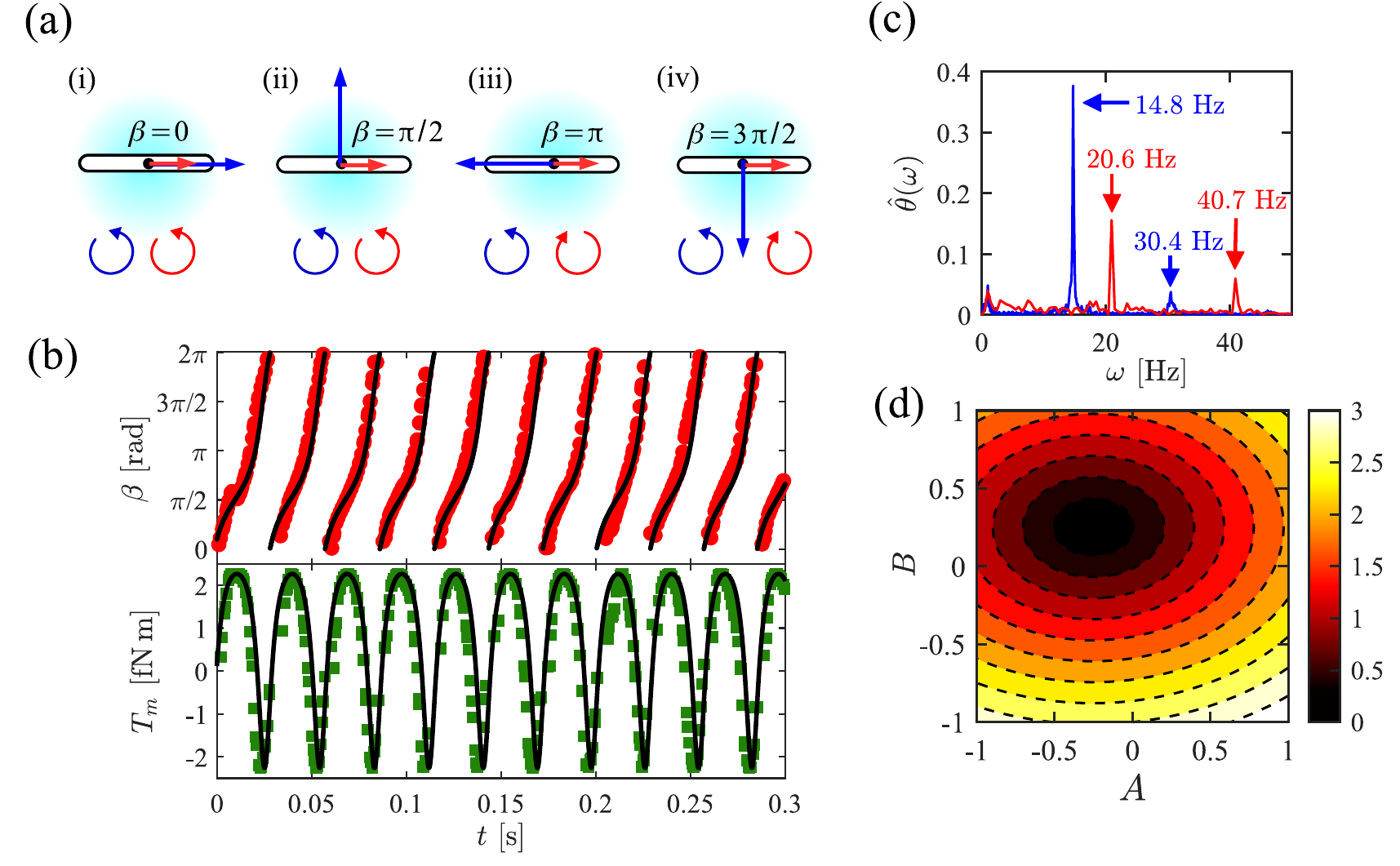}
	\caption{(a) Schematic illustration describing the behavior of rods phase angle $\beta$ beyond the step-out frequency in the viscoelastic case corresponding to the extreme values of magnetic torque $T_m$. (b) Steady state behavior of the phase angle $\beta$ and the resulting magnetic torque $T_m$, measured at $\mathbf{|B|}=30~\mathrm{G}$ and $\omega_B=40~\mathrm{Hz}$ in experiments (symbols) and in numerical simulations (solid curves). (c) Fourier transformed angular coordinate $\hat{\theta}(\omega)$ for $\omega_B=15~\mathrm{Hz}$ (blue curve) and $\omega_B=20~\mathrm{Hz}$ (red curve) beyond step-out illustrating the two dominating frequencies corresponding to $\omega_B$ and $2\omega_B$. (d) RMSE obtained from fitting Eqn.~\ref{eq:alternative_torque} to the experimentally measured $T_m$ for various values of $A$ and $B$, spanning from $-1$ to $+1$.}
\label{fig:panel2}
\end{figure*}

 \section{Underlying Mechanism}
To attain a comprehensive understanding of the rotational rod motion , in the viscoelastic case, particularly beyond the step-out regime ($\omega_B>\Omega_C$), we investigated the time evolution of the phase angle, $\beta(t)$ \textit{i.e.}, the angle between $\mathbf{m}$ and $\mathbf{B}$. Physically, for a given $\mathbf{|m|}$ and $\mathbf{|B|}$, $\beta(t)$ sets the magnitude of the instantaneous magnetic torque as $T_m(t)=\mathbf{|m|}\mathbf{|B|}\sin\beta$. Due to the sine dependency of $T_m$, it would bear extreme values $0,\pm\mathbf{|m||B|}$ when $\beta$ equals $0,\pi/2,\pi$ and $3\pi/2$ during a full cycle. A schematic depiction illustrating these four extreme cases are shown Fig.~\ref{fig:panel2}\,(a). A representative behavior of experimentally obtained $\beta(t)$ and $T_m(t)$ at $\omega_B=40~\mathrm{Hz}$ and $\mathbf{|B|}=30~\mathrm{G}$ is depicted in Fig.~\ref{fig:panel2}\,(b). Notably, $\beta(t)$ displays significant slope variation in comparison to the Newtonian case (Supplemental Material), during a complete $2\pi$ rotation. Particularly, it undergoes a dramatic reduction for $\pi/4\lesssim\beta\lesssim 3\pi/4$, which represents a part of phase-stick motion. Thereafter, it displays a rapid increase for $\beta\gtrsim3\pi/4$ which corresponds to the phase-slip motion. Interestingly, the phase-slip motion, which is expected to occur precisely at $\beta=\pi$ in a purely viscous scenario (See Supplemental Material), here, occurs for $\beta<\pi$, as evident by the heightened slope of $\beta$ before reaching $\pi$. What is even more surprising is that the rapid increase in $\beta$ is expected to cease at $\beta=0$ in the subsequent cycle where it aligns itself to the field, yet it persists until $\beta\approx\pi/4$. This strongly hints that the phase-slip motion is addditionally assisted by the relaxation mechanism of the viscoelastic fluid. \\

Intuitively, during the phase stick motion, the rotating rod induces a  strain within the viscoelastic network. The driving torque $T_m$ during this phase increases as $\beta$ evolves from $0\rightarrow\pi/2$, reaching its maximum value at $\beta = \pi/2$. For $\beta>\pi/2$, $T_m$ begins to decrease while at the same time the rod motion is strongly opposed by the previously strained viscoelastic fluid. This leads to a dramatic decrease in $\dot{\beta}(t)$ as observed $\beta>\pi/2$. Eventually, $T_m$ reaches a value where it is insufficient to move the rod against the strained viscoelastic fluid, resulting in a phase slip motion, where the relaxation process of the previously induced strain couples to the rod motion. This menifests as an abrupt increase of $\beta$. As discussed previously, this phenomenon occurred before $\beta$ reaches $\pi$ and persistes until $\beta\approx\pi/4$. The coupling, therefore, leads to considerably prolonged phase of stick motion when compared to the phase-slip motion as observed experimentally. This results in a notably enhanced net angular drift. Accordingly, $T_m$ exhibits a significant temporal asymmetry, with the positive phase of $T_m$ spanning a longer duration compared to its negative phase, as illustrated in Fig.~\ref{fig:panel2}\,(b).

\section{Minimal non-Markovian model}
As a matter of fact, the aforementioned surprising behavior of enhanced rod rotation can also be rationalized by means of a minimal non-Markovian model which takes into account the memory effects of the surrounding viscoelastic fluid. A stress-relaxation modulus which faithfully describes the mechanical response of such miceller solutions is given by the Jeffreys model $ G(t)=2\eta_\infty\delta(t)+[(\eta_0-\eta_\infty)/\tau]e^{-(t/\tau)}$~\cite{paul2018free,gomez2015transient}. Here, the first term corresponds to the bare solvent viscosity $\eta_\infty$ which responds instantaneously while the second term captures the slow microsturctural response of the fluid with stress-relaxation time $\tau$ and viscosity $\eta_0-\eta_\infty$. The mechanical response of the fluid couples to the rotational motion of the rod through the memory friction kernel $\Gamma_\theta(t)=\gamma_\theta G(t)$ where $\gamma_\theta$ is the geomertric friction factor of the rod which for a rod length $L$ and diameter $\sigma$ is $[3/(\pi L^3)] (\log(L/\sigma)-0.662+0.917(\sigma/L)-0.050(\sigma/L)^2)$~\cite{tirado1984comparison,narinder2022understanding}. Neglecting the inertial effects, we describe the rotational motion of a magnetic rod suspended in a homogeneous viscoelastic fluid under a rotating magnetic field $\mathbf{B}=|\mathbf{B}|\cos\omega_B t\,\hat{\mathbf{x}}+|\mathbf{B}|\sin\omega_B t\,\hat{\mathbf{y}}$ by the generalized Langevin equation
\begin{equation}
    -\int_{-\infty}^{t}\Gamma(t-t')\dot{\theta}(t')dt'+\mathbf{|m||B|}\sin\beta(t)+\xi_\theta(t)=0.
    \label{eq:gle}
\end{equation}
Here, $\xi_\theta(t)$ is zero mean Gaussian noise which mimics the thermal fluctuations of the fluid. The auto-correlation function of the noise obeys $\langle\xi_\theta(t)\xi_\theta(s)\rangle=k_B T\Gamma(t-s)$ which satisfies the fluctuation-dissipation theorem. By extending the degrees of freedom, Eqn.~\ref{eq:gle} can be recast to its Markovian equivalent~\cite{villamaina2009fluctuation,narinder2018memory}
\begin{equation}
\begin{bmatrix}
\dot{\theta}\\ \dot{\Theta}
\end{bmatrix}=\begin{bmatrix}
\frac{-\delta\eta}{\eta_\infty\tau} & \frac{\delta\eta}{\eta_\infty\tau}\\ 
 \frac{1}{\tau} & -\frac{1}{\tau}
\end{bmatrix}\begin{bmatrix}
\theta\\  \Theta
\end{bmatrix}+\begin{bmatrix}
\mathbf{|m||B|}\sin{\beta}\\ 
0
\end{bmatrix}+\begin{bmatrix}
\sqrt{2D^\infty_\theta}\phi\\ 
\sqrt{2D^{\delta\eta}_\theta}\varphi
\end{bmatrix}.
\label{eq:markoveq}
\end{equation}
Here, $\Theta(t)$ is an auxiliary variable for the angular coordinate $\theta(t)$, $D^{\infty}_\theta$ and $D^{\delta\eta}$ are the rotational diffusion coefficients of the rod about its long axis, related to the viscosities $\eta_\infty$ and $\delta\eta=\eta_0-\eta_\infty$, respectively. In addition, $\phi$ and $\varphi$ are two independent Gaussian white noises with zero mean \textit{i.e.}, $\langle\phi(t)\rangle =\langle\varphi(t)\rangle =0$ and autocorrelation $\langle\phi(t)\phi(s)\rangle=\langle\varphi(t)\varphi(s)\rangle =\delta(t-s)$.

%
% \begin{figure}[!]
%     \centering
%      \includegraphics[scale=0.6, trim={4cm 10.5cm 4cm 10cm}]{panel3.pdf}
%     \caption{(a) Numerically simulated time-evolution of the rod angle $\theta$ for various frequencies of applied $\mathbf{B}$. For the synchronous motion (straight curves), $\omega_B$ is increasing from bottom to top in the following order: $1~\mathrm{rad\,s^{-1}},~11~\mathrm{rad\,s^{-1}},~21~\mathrm{rad\,s^{-1}},~31~\mathrm{rad\,s^{-1}},~51~\mathrm{rad\,s^{-1}}$. For the asynchronous motion (oscillating curves), $\omega_B$ is increasing from top to bottom as $141~\mathrm{rad\,s^{-1}},~211~\mathrm{rad\,s^{-1}},~271~\mathrm{rad\,s^{-1}}$. (b) Numerically obtained angular velocity dependence on the applied frequency of the field at $B=30~\mathrm{G}$), normalized by the step-out value $\Omega_\mathrm{C}$, for the viscous parameters corresponding to $7~\mathrm{mM}$ CpyCl/NaSal solution for different values of the stress-relaxation time $\tau$. The dotted curve represents the corresponding Newtonian case.}
%    \label{fig:panel3}
% \end{figure}

The numerical solution of  Eqn.~\ref{eq:markoveq}, requires the knowledge of the rheological parameters of the fluid namely, $\eta_\infty$, $\eta_0$ and, $\tau$. These are obtained by passive microrheology via fitting the mean squared displacements $\langle\Delta r(t)^2\rangle$ of embedded spherical probes (see Supplemental Material). Interestingly, this minimal model captures all the salient features of the rotational rod dynamics in a viscoelastic fluid. In order to quantitatively compare the numerical results with the experiments, we plot the numerical data (solid curve) to the corresponding experimentally measured data in Fig.~\ref{fig:panel1}\,(d) where the magnetic moment $\mathbf{m}$ in numerical simulations is determined by matching the step-out value $\Omega_C$ to the experimental value. Opposed to a Newtonian liquid, here, the $\Omega$ beyond $\Omega_C$ depends strongly on the rheological parameters of the fluid. For the same viscous parameters \textit{i.e.}, $\eta_\infty$ and $\eta_0$, $\Omega$ beyond $\Omega_C$ drops slower with increasing $\tau$. Consistently, for vanishingly small $\tau$, the behavior converges towards the Newtonian case. Hence, confirms its generic nature which originates due to coupling of the stress-relaxation process of the viscoelastic fluid to the probe dynamics (Supplemental Material). Moreover, in Fig.~\ref{fig:panel2}\,(b), we show the numerically obtained $\beta(t)$ and $T_m(t)$ as solid curves along with our experimental results where an excellent agreement validate the underlying proposed physical mechanism for the net rotation. 
 \\

%\begin{figure}[!]
%	\centering
%	\includegraphics[scale=0.4]{underlying_mechanism.pdf}
%	\caption{(a) Numerically simulated time-evolution of the rod angle $\theta$ for various frequencies of applied $\mathbf{B}$. For the synchronous motion (straight curves), $\omega_B$ is increasing from bottom to top in the following order: (b) Numerically obtained angular velocity dependence on the applied frequency of the field at $B=30~\mathrm{G}$), normalized by the step-out value $\Omega_\mathrm{C}$, for the viscous parameters corresponding to $7~\mathrm{mM}$ CpyCl/NaSal solution for different values of the stress-relaxation time $\tau$. The dotted curve represents the corresponding Newtonian case.}
%	\label{fig:underlying_mechanism}
%\end{figure}

\section{Analytical solution of $\theta(t)$}
In the following, we seek an analytical solution of Eqn.~\ref{eq:gle} for $\theta(t)$ beyond step-out frequency in absence of thermal noise. This can be obtained by transforming Eqn.~\ref{eq:gle} into the frequency domain. Following this, the Laplace transformation of Eqn.~\ref{eq:gle} reads
\begin{equation}
 -s\tilde{\Gamma}(s)\tilde{\theta}(s)+\tilde{T}_m(s)=0 
 \label{eq:gle22}
 \end{equation}
However, the problem is that the time-domain expression of the magnetic torque $T_m(t)=\mathbf{|mB|}\sin(\omega_B t-\theta(t))$ depends on $\theta(t)$ which is not known \textit{a priori}. Therefore, its laplace-domain $\tilde{T}_m(s)$ can not be computed. We, thus, seek an approximate equivalent mathematical expression of $T_m(t)$, as an alternative, which provides us the same experimental results. For this purpose, we first note that the $T_m(t)$ is time-periodic with its maximum and minimum values being $+\mathbf{|mB|}$ and $-\mathbf{|mB|}$, respectively, as plotted in Fig.~\ref{fig:panel2}\,(b). Additionally, to obtain the dominant frequencies present in the torque, we transformed the torque to the frequency domain. A typical Fourier-transformed $\hat{\theta}(\omega)$ for $\omega_B=15~\mathrm{Hz}$ and $20~\mathrm{Hz}$ are plotted in Fig.~\ref{fig:panel2}\,(c). Vividly, $\hat{\theta}(\omega)$ is governed by the two dominant frequencies \textit{i.e.}, $\approx \omega_B$ and $\approx 2\omega_B$. From the physical point of view, it makes a perfect sense as during the phase-stick motion, the rod rotation follows the frequency of the applied field $\omega_B$ whereas during the phase-slip motion, it is additionally acted upon the micro-structure relaxation mechanism of the viscoelastic fluid. The fluid's response depends on the previously applied deformation rate which is $\omega_B$, as described in ref~\cite{gomez2015transient}. This combination (magnetic moment tries to align with the field along with the slow-microstructual relaxation) when acts in the same direction renders the $2\omega_B$ a dominant frequency in the phase-slip motion. From these hints, we write the alternative expression of the torque as 
\begin{equation}
T'_m(t)=\mathbf{|m||B|}\left ( \cos(\omega_B t) +A \cos(2\omega_B t)+B\right )
\label{eq:alternative_torque}
\end{equation}
where $A$ and $B$ are obtained to $-0.25$ and $0.25$, respectively, by minimizing the RMSE value with respect to the experimental $T_m$ as shown in Fig.~\ref{fig:panel2}\,(d).
Substituting the laplace-transformed $\Gamma(t)$ and $T_m'(t)$ in Eqn.~\ref{eq:gle22}, it reads
\begin{equation}
\tilde\theta(s)=\frac{mB(1+\tau s)(s^4+5s^2{\omega_B}^2+\omega^4)}{\gamma s^2 (s^4+5s^2{\omega_B}^2+4{\omega_B}^4) (\eta_0+\eta_\infty\tau s)}.  
\label{eq:theta_freq}
\end{equation}
Transforming Eqn.~\ref{eq:theta_freq} back to the time domain gives
% \begin{dmath}
% \begin{widetext}
% \begin{equation}
% \theta (t)= \frac{mB}{\gamma} \lvert \left (
%  \frac{\tau^2\eta_{\infty}\omega^2\sin(\omega t)+\tau\omega(\eta_0-\eta_\infty)\cos(\omega t)+\eta_0\sin(\omega t)}{\omega (\tau^2\eta_\infty^2\omega^2+\eta_0^2)}\right )
%  +\left ( \frac{-4\tau^2\eta_\infty\omega^2\sin(2\omega t)+2\tau(\eta_0-\eta_\infty)\omega\cos(2\omega t)-\eta_0\sin(2\omega t)}{8\omega(4\tau^2\eta_\infty^2\omega^2+\eta_0^2)} \right )
%    + e^{-\frac{\eta_0 t} {\eta_\infty\tau}} \frac{\tau^6\eta_\infty^6\omega^4-\tau^6\eta_\infty^5\eta_0\omega^4+5\tau^4\eta_\infty^4\eta_0^2\omega^2-5\tau^4\eta_\infty^3\eta_0^3\omega^2
%  +\tau^2\eta_\infty^2\eta_0^4-\tau^2\eta_\infty\eta_0^5}{\tau \eta_\infty \eta_0^2(\tau^2\eta_\infty^2\omega^2+\eta_0^2)(4\tau^2\eta_\infty^2\omega^2+\eta_0^2)} 
%  +\left ( \frac{\tau(\eta_0-\eta_\infty)}{4\eta_0^2} \right )+\left (\frac{t}{4\eta_0}\right) \rvert 
% \end{equation}
% \end{widetext}
% \end{dmath}

\begin{widetext}
\begin{equation}
\theta(t) = \frac{mB}{\gamma} \left|
\begin{aligned}
& \frac{\tau^2\eta_{\infty}\omega^2\sin(\omega t)}{\omega (\tau^2\eta_\infty^2\omega^2+\eta_0^2)} 
  + \frac{\tau\omega(\eta_0-\eta_\infty)\cos(\omega t)}{\omega (\tau^2\eta_\infty^2\omega^2+\eta_0^2)} 
  + \frac{\eta_0\sin(\omega t)}{\omega (\tau^2\eta_\infty^2\omega^2+\eta_0^2)} \\
& \quad - \frac{4\tau^2\eta_\infty\omega^2\sin(2\omega t)}{8\omega(4\tau^2\eta_\infty^2\omega^2+\eta_0^2)} 
  + \frac{2\tau(\eta_0-\eta_\infty)\omega\cos(2\omega t)}{8\omega(4\tau^2\eta_\infty^2\omega^2+\eta_0^2)} 
  - \frac{\eta_0\sin(2\omega t)}{8\omega(4\tau^2\eta_\infty^2\omega^2+\eta_0^2)} \\
& \quad + e^{-\frac{\eta_0 t}{\eta_\infty\tau}} \left(
    \frac{\tau^6\eta_\infty^6\omega^4 - \tau^6\eta_\infty^5\eta_0\omega^4 + 5\tau^4\eta_\infty^4\eta_0^2\omega^2 - 5\tau^4\eta_\infty^3\eta_0^3\omega^2}
    {\tau \eta_\infty \eta_0^2(\tau^2\eta_\infty^2\omega^2+\eta_0^2)(4\tau^2\eta_\infty^2\omega^2+\eta_0^2)} 
    + \frac{\tau^2\eta_\infty^2\eta_0^4 - \tau^2\eta_\infty\eta_0^5}
    {\tau \eta_\infty \eta_0^2(\tau^2\eta_\infty^2\omega^2+\eta_0^2)(4\tau^2\eta_\infty^2\omega^2+\eta_0^2)} \right) \\
& \quad + \left(\frac{\tau(\eta_0-\eta_\infty)}{4\eta_0^2} \right) 
  + \left(\frac{t}{4\eta_0}\right)
\end{aligned}
\right|
\label{eq:time_domain_theta}
\end{equation}
\end{widetext}

% \begin{equation} \label{eq:theta_time_domain}
% \begin{split}
% \theta (t) =\frac{m B}{\gamma} \Bigg| 
%     \left( 
%         \frac{\tau^2\eta_{\infty}{\omega_B}^2\sin({\omega_B} t) + \tau{\omega_B}
%         (\eta_0-\eta_\infty)\cos({\omega_B} t) + \eta_0\sin({\omega_B} t)}
%         {{\omega_B} (\tau^2\eta_\infty^2{\omega_B}^2 + \eta_0^2)}
%     \right) \\
%    &  + \left( 
% \frac{-4\tau^2\eta_\infty{\omega_B}^2\sin(2{\omega_B} t) + 2\tau(\eta_0-\eta_\infty){\omega_B}\cos(2\textbf{\omega} t) - \eta_0\sin(2{\omega_B} t)}
%         {8{\omega_B}(4\tau^2\eta_\infty^2{\omega_B}^2 + \eta_0^2)}\right) \\
%     & + e^{-\frac{\eta_0 t}{\eta_\infty\tau}} \frac{\tau^6\eta_\infty^6{\omega_B}^4 - \tau^6\eta_\infty^5\eta_0{\omega_B}^4 + 5\tau^4\eta_\infty^4\eta_0^2{\omega_B}^2 - 5\tau^4\eta_\infty^3\eta_0^3{\omega_B}^2
%         + \tau^2\eta_\infty^2\eta_0^4 - \tau^2\eta_\infty\eta_0^5}
%         {\tau \eta_\infty \eta_0^2(\tau^2\eta_\infty^2{\omega_B}^2 + \eta_0^2)(4\tau^2\eta_\infty^2{\omega_B}^2 + \eta_0^2)} \\
%     & + \left( \frac{\tau(\eta_0 - \eta_\infty)}{4\eta_0^2} \right) + \left( \frac{t}{4\eta_0} \right) \Bigg|
% \label{eq:theta_time_domain}
% \end{split}
% \end{equation}

\begin{figure}[!]
    \centering
     \includegraphics[scale=0.46]{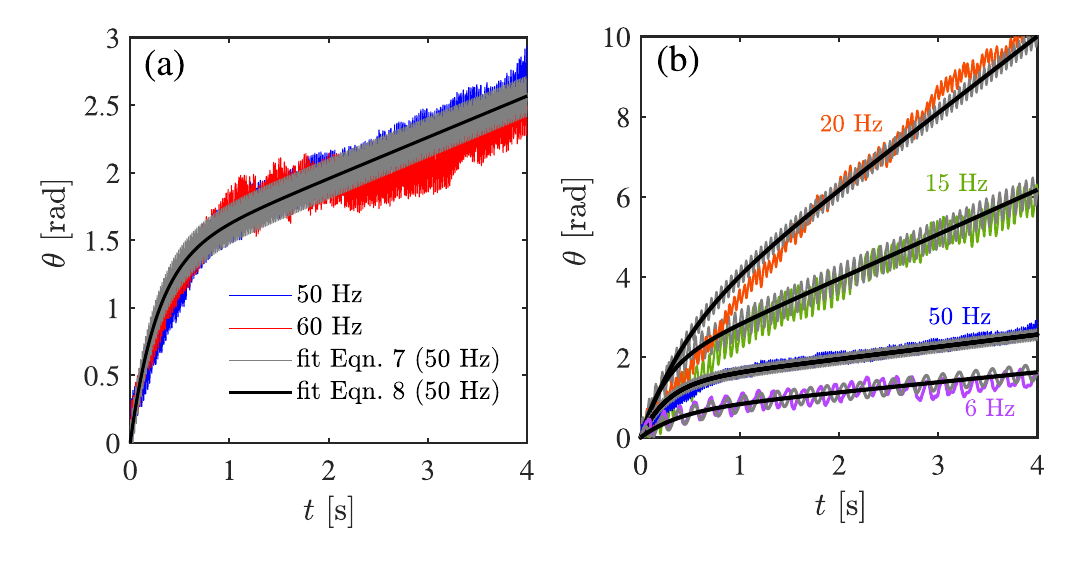}
     \caption{(a) Comparison of the experimentally measured time-evolution of the rod angular coordinate $\theta(t)$ in the CPyCl/NaSal viscoelastic solution for applied $\omega_B=50~\mathrm{Hz}$ and $60~\mathrm{Hz}$ at $10~\mathrm{G}$ and the analytical results Eqn.~\ref{eq:time_domain_theta} and Eqn.~\ref{eq:limit_high_freq}. (b) Evaluation of the experimentally obtained angular coordinate $\theta(t)$ and the analytical results Eqn.~\ref{eq:time_domain_theta} and \ref{eq:limit_high_freq} for different rods with varying magnetic moments. The analysis is conducted at an applied magnetic field of $|\mathbf{B}|=7~\mathrm{G}$ for the bottom-most curve and $|\mathbf{B}|=10~\mathrm{G}$ for the remaining curves, across various values of $\omega_B$.}
     \label{fig:high_freq}
     \end{figure}

Interestinlgy, Eqn.~\ref{eq:time_domain_theta} predicts the long time slope of $\theta(t)$, for a rod with magnetic moment $|\mathbf{m}|$ and applied $|\mathbf{B}|$ depends only by the zero-shear viscosity $\eta_0$ of the viscoelastic fluid. Indeed, we find that this prediction agrees with our experimental observation as shown in Fig.~\ref{fig:high_freq}\,(a), for a rod at $\omega_B=50~\mathrm{Hz}$ and $60~\mathrm{Hz}$ where Eqn.~\ref{eq:time_domain_theta}(gray curve) quantitatively captures both the short and long time behavior of experimentally measured $\theta(t)$.
Additionally, for the limiting case $\omega_B\rightarrow \infty$, Eqn.~\ref{fig:high_freq} can be simplified as
\begin{equation}
\theta(t) = \frac{mB}{4\eta_0^2\gamma} \left( \tau(\eta_0 - \eta_{\infty}) \left(1 - e^{-\frac{\eta_0 t}{\eta_{\infty} \tau}}\right) + \eta_0 t \right).
\label{eq:limit_high_freq}
\end{equation} 
Eqn.~\ref{eq:limit_high_freq} describes the mean observed dynamics in a minimal manner, as shown by the black solid curve in Fig.~\ref{fig:high_freq}\,(a). This can be exploited straightforwardly to obtain the rheological parameters of a fluid.\\ 
In Fig.~\ref{fig:high_freq}\,(b), we compared the analytical result of Eqn.~\ref{eq:time_domain_theta} for rods with different $|\mathbf{m}|$ and applied $|\mathbf{B}|$ and different $\omega_B$ against the experimentally obtained $\theta(t)$.  Notably, the theoretical predictions exhibit excellent agreement with the experimental data under various conditions. 

\section{Conclusion}
In conclusion, we studied the rotational motion of colloidal rod-shaped particles in a viscoelastic fluid under the influence of a rotating magnetic field across a wide frequency range. Our findings revealed a notable disparity of rotational rod dynamics in a viscoelastic fluid beyond the step-out frequency compared to a Newtonian fluid. Even at frequencies where the rod ceases to rotate in a purely viscous fluid, it exhibits a significant angular frequency in a viscoelastic case. Furthermore, we demonstrated that this distinctive characteristic of rod motion is not specific to the investigated viscoelastic fluid but also applies to other types of viscoelastic media~\cite{hunter2012physics,zaccarelli2009colloidal,moeendarbary2013cytoplasm}. We proposed that the observed behavior emerges from the coupling of stress-relaxation process of the viscoelastic fluid to the rotational rod motion. Our explaination is further corrobated by a minimal non-Markovian model that incorporates the memory effects of the surrounding viscoelastic environment into the rotational motion of the rod. This coupling leads to a large variation in the magnitudes of the stick and slip angular motion, which promotes a net forward rotation. Our findings provide a deeper understanding of the temporal information of driven Brownian particles in viscoelastic media under transient stresses which is usually obsecured in coarse-grained descriptions. Our findings will have dramatic consequences for understanding the dynamics probes \textit{e.g.}, with additional interlinked degrees of freedom~\cite{spagnolie2013locomotion,kummel2013circular}, in field gradients~\cite{valberg1987magnetic} and in confinements~\cite{narinder2019active}. Furthermore, they would have significant implications for elucidating critical processes in biological systems, where intricate interactions between active fluctuations and viscoelastic environments are at play~\cite{joanny2009active,almonacid2015active}.

\begin{acknowledgments}
N.N. acknowledges support from the IISc IoE post-doctorate fellowship. 
\end{acknowledgments}

%\bibliography{references}% Produces the bibliography via BibTeX.

\end{document}